\documentclass[prd,onecolumn,amsmath,amssymb,floatfix,nofootinbib]{revtex4}

\usepackage{amssymb}
\usepackage{amsmath}
\usepackage{slashed}

\usepackage{caption}
\usepackage{subfigure}

\usepackage{graphicx}
\usepackage{graphics}
\usepackage{dcolumn}
\usepackage{color}
\usepackage{rotate}
\usepackage{fancyhdr}
\usepackage{hyperref}
\hypersetup{
    colorlinks=true,
    linkcolor=black,
    filecolor=black,
    urlcolor=black,
}
\usepackage{indentfirst}
\usepackage[titletoc,title]{appendix}

\def\be{\begin{eqnarray}}
\def\ee{\end{eqnarray}}
\def\no{\nonumber}

\definecolor{darkred}{rgb}{.743,0,0}

\def\lsim{\mathrel{\rlap{\lower4pt\hbox{\hskip1pt$\sim$}}
     \raise1pt\hbox{$<$}}}         
\def\gsim{\mathrel{\rlap{\lower4pt\hbox{\hskip1pt$\sim$}}
     \raise1pt\hbox{$>$}}}         
\newcommand{\beq}{\begin{equation}}
\newcommand{\eeq}{\end{equation}}

\newenvironment{Eqnarray}{\arraycolsep 0.14em\begin{eqnarray}}{\end{eqnarray}}
\def\beqa{\begin{Eqnarray}}
\def\eeqa{\end{Eqnarray}}

\begin{document}
\title{Testing Minimal Flavor Violation in Leptoquark Models of the $R_{K^{(*)}}$ Anomaly}

\author{Daniel Aloni$^{1a}$, Avital Dery$^{1a}$, Claudia Frugiuele$^{1a}$ and Yosef Nir$^{1a}$}
\affiliation{$^1$Department of Particle Physics and Astrophysics, Weizmann Institute of Science, Rehovot, Israel 7610001}
\email{$^a$daniel.aloni, avital.dery, claudia.frugiuele, yosef.nir@weizmann.ac.il}

\begin{abstract}
\noindent
The $R_{K^{(*)}}$ anomaly can be explained by tree level exchange of leptoquarks. We study the consequences of subjecting these models to the principle of minimal flavor violation (MFV). We consider MFV in the linear regime, and take the charged lepton Yukawa matrix to be the only spurion that violates lepton flavor universality. We find that a combination of constraints from a variety of processes -- $b\to s\mu\mu$, $b\to s\tau\tau$, $b\to s\nu\nu$, $b\bar b\to\tau\tau$ and  $b\to c\tau\nu$ -- excludes MFV in these models.
\end{abstract}

\maketitle


\vskip 8pt

%
%

\vglue 0.3truecm


\section{Introduction}
Within the Standard Model (SM), lepton flavor universality (LFU) is respected by the weak interactions. Consequently, LFU is predicted to hold -- up to (calculable) phase-space effects -- in processes where the Yukawa interactions are negligible. Hints of violation of LFU have, however, been observed by the LHCb experiment in $B\to K^{(*)}\ell^+\ell^-$ decays. While LFU implies that the ratios
\beq
R_{K^{(*)},[a,b]}=\frac{\int_a^b dq^2[d\Gamma(B\to K^{(*)}\mu^+\mu^-)/dq^2]}{\int_a^b dq^2[d\Gamma(B\to K^{(*)}e^+e^-)/dq^2]}
\eeq
($q^2$ is the invariant dilepton mass-squared) should be very close to unity, the measurements give \cite{Aaij:2014ora,Aaij:2017vbb}
\beqa
R_{K,[1,6]{\rm GeV}^2}&=&0.745^{+0.090}_{-0.074}\pm0.036,\no\\
R_{K^*,[1.1,6.0]{\rm GeV}^2}&=&0.69^{+0.11}_{-0.07}\pm0.05,\no\\
R_{K^*,[0.045,1.1]{\rm GeV}^2}&=&0.66^{+0.11}_{-0.07}\pm0.03,
\eeqa
which stand in a $2.2-2.6\sigma$ discrepancy with the SM predictions.

The discrepancy, if not a statistical fluctuation, requires new degrees of freedom. In this work we focus on new physics models where heavy new bosons contribute to $b\to s\mu^+\mu^-$ transitions at tree level. Such new bosons can be $SU(3)_C$ singlets or triplets. We focus on the latter class, {\it i.e.} on leptoquark models \cite{Hiller:2014yaa}. More specifically, we consider simplified models where a single leptoquark representation is added to the SM fields.

There are eight leptoquark representations that have couplings to down-type quarks and to charged leptons. The $R_{K^{(*)}}$ measurements suggest that the integration out of these leptoquarks generate an effective four-fermi operator of the form
\beq\label{eq:bsmumu}
C_{bs\mu\mu}(\overline{s_L}\gamma^\mu b_L)(\overline{\mu_L}\gamma_\mu\mu_L).
\eeq
Accordingly, the eight leptoquark representations can be divided to three groups:
\begin{itemize}
\item One of the scalar leptoquark representations,
\beq
S(3,1)_{-1/3},
\eeq
couples down quarks to neutrinos and up quarks to the charged leptons, so it does not generate (at tree level) the operator of Eq. (\ref{eq:bsmumu}).
\item The couplings of three of the scalar and one of the vector leptoquark representations,
\beq
S^\prime(3,1)_{-4/3},\ \ \ D(3,2)_{+7/6},\ \ \ D^\prime(3,2)_{+1/6},\ \ \ V^\mu(3,2)_{-5/6},
\eeq
involve right-handed fields and thus cannot explain the anomaly.
\item The remaining one scalar and two vector leptoquark representations,
\beq\label{eq:viablelq}
T(3,3)_{-1/3},\ \ \ U_3^\mu(3,3)_{+2/3},\ \ \ U_1^\mu(3,1)_{+2/3},
\eeq
are viable candidates to explain the $R_{K^{(*)}}$ anomaly \cite{Hiller:2017bzc,Alonso:2015sja,Becirevic:2016oho,Crivellin:2017zlb,Alok:2017sui,Fajfer:2015ycq}.
\end{itemize}
In this work, we thus focus on the three simplified models of Eq. (\ref{eq:viablelq}).

The requirement that the contribution of the leptoquarks to $R_{K^{(*)}}$ breaks LFU implies that the leptoquark couplings have a non-trivial flavor structure. In particular, they must break the accidental $SU(3)_Q\times SU(3)_L$ global symmetry of the gauge interactions. Generic breaking would lead to unacceptably large contributions to various flavor changing processes. This situation is the specific realization of the new physics flavor puzzle \cite{Nir:2013maa} in the leptoquark framework. Thus, the $R_{K^{(*)}}$ measurements provide an opportunity to test the various ideas that have been proposed to solve this puzzle \cite{Alonso:2015sja,Calibbi:2015kma,Buttazzo:2017ixm,Cline:2017ihf}. Arguably the simplest, and the most easily falsifiable of these is the principle of minimal flavor violation (MFV) \cite{DAmbrosio:2002vsn}. In this work we ask whether the leptoquark models that explain the $R_{K^{(*)}}$ anomaly can be MFV (see \cite{Alonso:2015sja} for related work).

Within the MFV framework, various flavor changing processes are related to each other. For example, the $b\to s\mu^+\mu^-$ transition relevant to $R_{K^{(*)}}$ is related to the $b\to s\tau^+\tau^-$ and the $b\to d\mu^+\mu^-$ transitions. We ask whether the MFV relations exclude some or all of the three otherwise-viable leptoquark models.

The plan of this paper goes as follows. In Section \ref{sec:mfvlq} we present the principles of applying MFV on leptoquark couplings. In Section \ref{sec:mlfv} we obtain the viable lepton flavor representations for leptoquarks, and exclude some of the gauge representations that would be viable if MFV were not imposed. In Section \ref{sec:mqfv} we test the various quark flavor representations against experimental constraints. We present our conclusions in Section \ref{sec:conclusions}. Several additional phenomenological constraints are discussed in Appendices: $B_s-\overline{B_s}$ mixing (Appendix \ref{sec:bsubs}), direct LHC searches (Appendix \ref{sec:directLHC}), perturbative unitarity (Appendix \ref{sec:direct}), $pp\to\mu\mu$ (Appendix \ref{sec:ppmumu}), and $s\to u\tau\nu$ (Appendix \ref{sec:rtauk}).

\section{MFV for leptoquarks}
\label{sec:mfvlq}
In this section we discuss in more detail the implementation of MFV in leptoquark models \cite{Davidson:2010uu}. In the absence of Yukawa couplings, the SM acquires an accidental non-Abelian global symmetry,
\beqa
G_{\rm flavor}&=&SU(3)_q^3\times SU(3)_\ell^2,\\
SU(3)_q^3&=&SU(3)_Q\times SU(3)_U\times SU(3)_D,\no\\
SU(3)_\ell^2&=&SU(3)_L\times SU(3)_E.\no
\eeqa
The Yukawa couplings,
\beq
{\cal L}_{\rm Yukawa}=\overline{Q}Y^D D\phi+\overline{Q}Y^U U\widetilde\phi+\overline{L}Y^E E\phi+{\rm h.c.},
\eeq
break $G_{\rm flavor}\to U(1)_B\times U(1)_e\times U(1)_\mu\times U(1)_\tau$. Thus, the three Yukawa matrices can be taken as spurions with the following transformation properties under $G_{\rm flavor}$:
\beq
Y^U(3,\bar3,1,1,1),\ \ \ Y^D(3,1,\bar3,1,1),\ \ \ Y^E(1,1,1,3,\bar3).
\eeq
Imposing MFV on the SM extended with leptoquark fields means that we assign the leptoquark fields with well-defined transformation properties under $G_{\rm flavor}$ and require the following:
\begin{itemize}
\item All terms made of SM fields, leptoquark fields and the Yukawa spurions are formally invariant under $G_{\rm flavor}$.
\end{itemize}
One subtlety relates to the definition of minimal lepton flavor violation. We consider the case that the only spurion that breaks $SU(3)^2_\ell$ is $Y^E$. If one takes into account the fact that neutrinos are massive, additional spurions may play a role. For example, if neutrino masses arise from a seesaw mechanism with three heavy SM-singlet fermions $N$, then $G_{\rm flavor}$ is extended by an $SU(3)_N$ factor, and both $M_N$, the mass matrix of these fermions, and $Y^N$, the neutrino Yukawa matrix, break the flavor symmetry. Taking $Y^E$ to be the only leptonic spurion is equivalent to assuming that the seesaw scale is higher than the scale at which the leptoquark couplings are set. Moreover, if this scenario holds in Nature, it explains why lepton flavor violation ({\it e.g.}, $\mu\to e\gamma$ \cite{Crivellin:2017dsk}) has not been observed except in neutrino oscillations.

We are interested in leptoquarks that generate the effective four-fermi operator of Eq. (\ref{eq:bsmumu}). Thus, the $SU(3)_C\times SU(2)_L\times U(1)_Y$ invariant operator must involve the leptoquark field, the quark doublet fields $Q_i$ and the lepton doublet fields $L_j$. Since our starting point is the anomaly in $b\to s\mu^+\mu^-$ transitions, we work in the down and charged lepton mass basis. Hence the quark doublets are $Q_{d,s,b}$ and the lepton doublets are $L_{e,\mu,\tau}$. In this basis, the three Yukawa spurions have the form
\beqa
Y^D&=&\lambda_d\equiv{\rm diag}(y_d,y_s,y_b),\no\\
Y^U&=&V^\dagger\lambda_u\equiv V_{\rm CKM}^\dagger\times{\rm diag}(y_u,y_c,y_t),\no\\
Y^E&=&\lambda_e\equiv{\rm diag}(y_e,y_\mu,y_\tau).
\eeqa

To have a predictive framework for processes that involve the third generation fermions (in particular the $b$-quark and the $\tau$-lepton), we make two assumptions:
\begin{enumerate}
\item The spurions related to $Y^D$ and $Y^E$ are small enough to keep the leptoquark couplings perturbative.
\item Terms that are higher power in $Y^F$ ($F=U,D,E$) are suppressed compared to lower powers.
\end{enumerate}
The first assumption can be satisfied in the models that we consider for leptoquark masses not much heavier than a few TeV. A quantitative analysis is given in Appendix \ref{sec:direct}.

The second assumption means that we do not consider MFV in the nonlinear regime \cite{Kagan:2009bn}. The implications of relaxing this assumption are briefly discussed in Section \ref{sec:conclusions}. The only case where we include spurions that are quadratic (or higher order) in the Yukawa couplings is when the leading contribution to flavor changing couplings arises from the operator
\beq
O^{Q_U}(8,1,1,1,1)\equiv Y^U Y^{U\dagger}.
\eeq
In the down mass basis, and neglecting $y_c$ and $y_u$, it has the form
\beq
(O^{Q_U})_{ji}=y_t^2 V_{tj}^*V_{ti}.
\eeq
%

\section{Minimal lepton flavor violation (MLFV)}
\label{sec:mlfv}
For the sake of concreteness we continue by considering a specific model out of the three -- that is the $T(3,3)_{-1/3}$ model -- but at this stage the lessons drawn are common to all three. The leptoquark couplings of $T$ have the form
\beq
{\cal L}_{\rm Yukawa}^T= \lambda_{\alpha j}\bar{Q}^c_j \epsilon (T_a^\dagger \tau_a)L_\alpha
+{\rm h.c.},
\eeq
where $\epsilon=i\tau_2$, and $\tau_a$ are Pauli matrices in $SU(2)_L$. Integrating out $T$, we obtain the following EFT Lagrangian:
\beq
{\cal L}^{\rm EFT}_T
= \frac{\lambda_{\alpha j}\lambda^{*}_{\beta i}}{M_T^2} (\overline{L_\beta}\tau_a \epsilon^T Q^c_i)(\overline{Q^c_j}\epsilon \tau_a L_\alpha).
\eeq
%

\subsection{$B\to K^{(*)}\ell^+\ell^-$}
The relevant leptoquark models generate, among others, operators of the following form:
\beq
C_{bs\mu\mu}^{\rm NP}(\overline{s_L}\gamma^\mu b_L)(\overline{\mu_L}\gamma_\mu\mu_L)+
C_{bs\tau\tau}^{\rm NP}(\overline{s_L}\gamma^\mu b_L)(\overline{\tau_L}\gamma_\mu\tau_L),
\eeq
with
\beq
C_{bs\ell\ell}^{\rm NP}=\frac{\lambda_{\ell b}^*\lambda_{\ell s}}{M_T^2}.
\eeq
We consider the experimental data, ${\rm BR}(B^+\to K^+\tau^+\tau^-)<2.25\times10^{-3}$ \cite{TheBaBar:2016xwe}, and ${\rm BR}(B^+\to K^+\mu^+\mu^-)=(4.4\pm0.3)\times10^{-7}$ \cite{Olive:2016xmw}, which give
\beq\label{eq:taumuexp}
R_{\tau/\mu}\equiv\frac{{\rm BR}(B^+\to K^+\tau^+\tau^-)}{{\rm BR}(B^+\to K^+\mu^+\mu^-)}<5\times10^3.
\eeq

We now examine various possibilities for the representation of $T$ under $SU(3)_\ell^2$ and their predictions for $\lambda_{\alpha i}$ and, consequently, for $C_{bs\mu\mu}$ and $C_{bs\tau\tau}$.

\begin{itemize}
\item $SU(3)_\ell^2$-singlet:
\end{itemize}
\beq
T(1,1)_{SU(3)_\ell^2}\ \Longrightarrow\ \lambda=0.
\eeq
The reason is that no combination of $Y^E$'s transforms as $(\bar3,1)_{SU(3)_\ell^2}$. Thus this MFV model cannot account for the $R_{K^{(*)}}$ anomaly.
\begin{itemize}
\item $SU(3)_L$-anti-triplet:
\end{itemize}
The spurion must transform as $(1+8,1)_{SU(3)_\ell^2}$ and thus
\beq
T(\bar3,1)\ \Longrightarrow\ \lambda\propto (1+Y^EY^{E\dagger}).
\eeq
Given the smallness of the lepton Yukawa couplings, we expect that the leading contribution is lepton-flavor universal and thus cannot account for the $R_{K^{(*)}}$ anomaly.

It could, however, be that the singlet contribution is negligibly small for some reason, and the octet contribution dominates. In the case of octet-spurion dominance, $\lambda\propto Y^EY^{E\dagger}$, we have
\beq\label{eq:octet}
\frac{C_{bs\tau\tau}^{\rm NP}}{C_{bs\mu\mu}^{\rm NP}}=\frac{y_\tau^4}{y_\mu^4}=8\times10^4.
\eeq
Taking into account that the ${\cal O}(0.25)$ deviation of $R_K$ from unity comes from the interference of the SM and leptoquark amplitudes, we find that Eq. (\ref{eq:octet}) implies $R_{\tau/\mu}\sim10^8$, strongly violating the experimental upper bound of Eq. (\ref{eq:taumuexp}). We conclude that having a leptoquark transform as $(\bar3,1)$ under $SU(3)_L\times SU(3)_E$ is excluded.
\begin{itemize}
\item $SU(3)_E$-anti-triplet:
\end{itemize}
\beq
T(1,\bar3)_{SU(3)_\ell^2}\ \Longrightarrow\ \lambda\propto Y^{E\dagger}.
\eeq
For all such models, we have the ratio between the $T$-mediated amplitudes given by
\beq\label{eq:triplet}
\frac{C_{bs\tau\tau}^{\rm NP}}{C_{bs\mu\mu}^{\rm NP}}=\frac{y_\tau^2}{y_\mu^2}=2.8\times10^2.
\eeq
Thus, these models predict
\beq
R_{\tau/\mu}\sim1.2\times10^3,
\eeq
a factor of 4 below the present bound.

\subsection{$B\to K^{(*)}\nu\bar\nu$}\label{sec:rnumu}
In the previous subsection, we proved that the only viable lepton flavor representation is $(1,\bar3)_{SU(3)_\ell^2}$. In this subsection we use the experimental data on $B\to K^{(*)}\nu\bar\nu$ to exclude some of these models.

Experiments put the upper bounds ${\rm BR}(B^+\to K^+\nu\bar\nu)<1.6\times10^{-5}$ \cite{Lees:2013kla} and ${\rm BR}(B^+\to K^{*+}\nu\bar\nu)<4.0\times10^{-5}$ \cite{Lutz:2013ftz}. Thus,
\beq\label{eq:rnumu}
R_{\nu/\mu}^{(*)}\equiv\frac{{\rm BR}(B^+\to K^{(*)+}\nu\bar\nu)}{{\rm BR}(B^+\to K^{(*)+}\mu^+\mu^-)}\lsim40.
\eeq

The relevant leptoquark models generate, among others, operators of the following form:
\beq
C_{bs\mu\mu}(\overline{s_L}\gamma^\mu b_L)(\overline{\mu_L}\gamma_\mu\mu_L)+
C_{bs\nu_\tau\nu_\tau}(\overline{s_L}\gamma^\mu b_L)(\overline{\nu_\tau}\gamma_\mu\nu_\tau).
\eeq
The SM predicts \cite{Buras:2014fpa,Geng:2017svp}
\beq
C_{bs\nu\nu}^{\rm SM}/C_{bs\mu\mu}^{\rm SM}=-1.49.
\eeq
%
%
%
(Note that $C_{bs\nu\nu}^{\rm SM}$ is the value for a single flavor, and thus the SM prediction is $R_{\nu/\mu}\sim6.6$.)
The $R_{K^{(*)}}$ anomaly requires
\beq
C_{bs\mu\mu}^{\rm NP}/C_{bs\mu\mu}^{\rm SM}=-0.12.
\eeq
We now obtain the ratio $C_{bs\nu_\tau\nu_\tau}^{\rm NP}/C_{bs\mu\mu}^{\rm NP}$ for each of $T$, $U_3^\mu$ and $U_1^\mu$, and the resulting prediction for $R_{\nu/\mu}$:
\beq
R_{\nu/\mu}\sim \frac{2|C_{bs\nu\nu}^{\rm SM}|^2+|C_{bs\nu_\tau\nu_\tau}^{\rm NP}+C_{bs\nu\nu}^{\rm SM}|^2}
{|C_{bs\mu\mu}^{\rm NP}+C_{bs\mu\mu}^{\rm SM}|^2}.
\eeq
Note that for $R_{\tau/\mu}$ we had to consider only the lepton flavor representation of the leptoquark. In contrast, for $R_{\nu/\mu}$, the result depends also on the Lorentz and $SU(2)\times U(1)$ representation and is thus different among the three models.
\begin{itemize}
\item $T(3,3)_{-1/3}$:
\beq
\frac{C_{bs\nu_\tau\nu_\tau}^T}{C_{bs\mu\mu}^T}=-\frac{y_\tau^2}{2y_\mu^2}\ \Longrightarrow\
R_{\nu/\mu}\sim3.1\times10^2.
\eeq
\item $U_3^\mu(3,3)_{+2/3}$:
\beq
\frac{C_{bs\nu_\tau\nu_\tau}^{U_3}}{C_{bs\mu\mu}^{U_3}}=-\frac{2y_\tau^2}{y_\mu^2}\ \Longrightarrow\
R_{\nu/\mu}\sim5.6\times10^3.
\eeq
\item $U_1^\mu(3,1)_{+2/3}$:
\beq
\frac{C_{bs\nu_\tau\nu_\tau}^{U_1}}{C_{bs\mu\mu}^{U_1}}=0\ \Longrightarrow\
R_{\nu/\mu}\sim8.5.
\eeq
\end{itemize}
We conclude that, for the $(1,\bar3)_{SU(3)_\ell^2}$ representation, the $T$ and $U_3^\mu$ models are excluded by the upper bound on $R_{\nu/\mu}$. On the other hand, the $U_1^\mu$ models predict this ratio to be a factor of 4.7 below the present bound (or, equivalently, 1.3 above the SM prediction).

\subsection{Summary of MLFV}
There are four classes of MLFV models for leptoquarks that can a-priori (that is, without imposing MLFV) generate the operator of Eq. (\ref{eq:bsmumu}):
\begin{itemize}
\item Models where it does not couple to the leptons. These are the models where the leptoquark transforms as $(1,1)_{SU(3)_\ell^2}$.
\item Models where the couplings are lepton-flavor-universal to a good approximation. This is the case if the leptoquark transforms as $(\bar3,1)_{SU(3)_\ell^2}$ and the leading spurion is a lepton flavor singlet.
\item Models where the leptoquark couplings are quadratic in the lepton-Yukawa. This is the case if the leptoquark transforms as $(\bar3,1)_{SU(3)_\ell^2}$ and the leading spurion is a lepton flavor octet.
\item Models where the leptoquark couplings are linear in the lepton-Yukawa. These is the case if the leptoquark transforms as $(1,\bar3)_{SU(3)_\ell^2}$.
\end{itemize}
Only the latter class is  good for explaining the $R_{K^{(*)}}$ anomaly (without violating the $R_{\tau/\mu}$ bound). This MFV classification is common to all three viable leptoquark models: $T(3,3)_{-1/3}$, $U_3^\mu(3,3)_{+2/3}$, and $U_1^\mu(3,1)_{+2/3}$. However, additional processes put further constraints:
\begin{itemize}
\item The upper bounds on ${\rm BR}(B\to K^{(*)}\nu\bar\nu)$ exclude the MFV-$T$ and MFV-$U_3^\mu$ models.
\item In Appendix \ref{sec:bsubs} we show that the MFV-$T$ model is excluded also by the upper bound on new physics contribution to $B_s-\overline{B_s}$ mixing.
\end{itemize}
We conclude that the only model that is not excluded by the above consideration is the $U_1^\mu$ model in the $(1,\bar3)_{SU(3)_\ell^2}$ representation.

To make further progress, we need to consider the $SU(3)_q^3$ representation of the leptoquark, which we do in the next section.

\section{Minimal Quark Flavor Violation (MQFV)}
\label{sec:mqfv}
We now consider the possible $SU(3)_q^3$ representations of the $U_1^\mu$ leptoquark. For simplicity, from here on we omit the sub-index 1 and the Lorentz super-index $\mu$ and denote the Lorentz-vector in the $(3,1)_{+2/3}$ gauge representation simply by $U$.
\begin{itemize}
\item $SU(3)_q^3$-singlet:
\end{itemize}
\beq
U(1,1,1)_{SU(3)_q^3}\ \Longrightarrow\ \lambda=0.
\eeq
The reason is that no combination of $Y^U$'s and $Y^D$'s transforms as $(3,1,1)_{SU(3)_q^3}$.

We conclude that $U$ must transform as a triplet under $SU(3)_q^3$ and as an anti-triplet under $SU(3)_\ell^2$. If indeed $U$ transforms as a quark--flavor-triplet and lepton-flavor-anti-triplet, then there are nine $U$-flavor states, that can be denoted as
\beq
U_{\alpha i},\ \ \alpha=e,\mu,\tau,\ \ \ i=d,s,b\ {\rm or}\ u,c,t.
\eeq

The $9\times9$ mass-squared matrix of the $U$ flavor states transforms as either $1$ or $1+8$ under each of the five $SU(3)$'s. Given the smallness of all Yukawa couplings except $y_t$, and the smallness of $|V_{ts}|$ and $|V_{td}|$, the $9\times9$ mass-squared matrix is near diagonal, so that we can call the nine $U$ mass-eigenstates by the same names as the flavor states, namely $U_{\alpha i}$. Furthermore, given our assumption of small $SU(3)_\ell^2$ spurions, the masses are lepton-flavor universal to a good approximation. As concerns the quark-flavor, in some cases the $b$-states ($t$-states) are separated by ${\cal O}(y_t^2)$ from the $s$- and $d$-states ($u$- and $c$-states), but in any case there is no hierarchy.

In what follows we denote the couplings of $U_{\alpha i}\overline{Q_j}L_\beta$ as $\lambda_{\alpha\beta ij}$:
\beq
{\cal L}_{\rm Yukawa}^U=\lambda_{\alpha\beta ij}\overline{Q_{j}}\gamma_\mu L_{\beta}U^\mu_{\alpha i}+{\rm h.c.}.
\eeq
As argued above, the only viable lepton flavor representation is $(1,\bar3)_{SU(3)_\ell^2}$ and thus
\beq
\lambda_{\alpha\beta ij}\propto(Y^{E\dagger})_{\alpha\beta}=\delta_{\alpha\beta}y_\beta,
\eeq
where the second equality applies in the charged lepton mass basis.

There are three possible $SU(3)_q^3$ representation. We denote the three models by $U_{Q,U,D}$ in correspondence to the flavor group -- $SU(3)_{Q,U,D}$ -- under which they transform as a triplet:
\begin{itemize}
\item $U_Q(3,1,1)_{SU(3)_q^3}$:\\
The required spurion transforms as $(1+8,1,1)_{SU(3)_q^3}$ and thus
\beq
\lambda_{\alpha\beta ij}\propto(x_{1/8}{\bf 1}+Y^U Y^{U\dagger})_{ji}=x_{1/8}\delta_{ji}+y_t^2V_{tj}^* V_{ti}.
\eeq
\item $U_U(1,3,1)_{SU(3)_q^3}$:\\
The required spurion transforms as $(3,\bar3,1)_{SU(3)_q^3}$ and thus
\beq
\lambda_{\alpha\beta ij}\propto Y^U_{ji}=V_{tj}^* y_t\delta_{it}.
\eeq
\item $U_D(1,1,3)_{SU(3)_q^3}$:\\
The required spurion transforms as $(3,1,\bar3)_{SU(3)_q^3}$ and thus
\beq
\lambda_{\alpha\beta ij}\propto[(x_{1/8}{\bf 1}+O^{Q_U})Y^D]_{ji}=y_b\delta_{ib}(x_{1/8}\delta_{jb}+y_t^2 V_{tj}^*V_{tb}).
\eeq
\end{itemize}
To summarize, we present the couplings that play a role in our framework in Table \ref{tab:couplings}.
\begin{table}[h!]
\caption{ \it The $\lambda_{\alpha\beta ij}$ couplings in the down and charged lepton mass basis for the three flavor representations.}
\label{tab:couplings}
\begin{center}
\begin{tabular}{ccc} \hline\hline
\rule{0pt}{1.0em}%
Model & $[SU(3)]^5$-rep & $\lambda_{\alpha\alpha ij}$  \\[2pt]
\hline\hline
$(U_Q)_{\alpha i}$ & $(3,1,1,1,\bar3)$ & $by_\alpha(x_{1/8}\delta_{ij}+y_t^2V_{tj}^*V_{ti})$ \rule{0pt}{1.0em}\\
$(U_U)_{\alpha i}$ & $(1,3,1,1,\bar3)$ & $b\delta_{it}y_\alpha y_tV_{tj}^*$ \rule{0pt}{1.0em}\\
$(U_D)_{\alpha i}$ & $(1,1,3,1,\bar3)$ & $b\delta_{ib}y_\alpha y_b(x_{1/8}\delta_{jb}+y_t^2V_{tj}^*V_{tb})$ \rule{0pt}{1.0em}\\
\hline\hline
\end{tabular}
\end{center}
\end{table}
%

\subsection{Back to $R_{K^{(*)}}$}
Given the couplings in Table \ref{tab:couplings}, we can now translate the $R_{K^{(*)}}$ requirement,
\beq\label{eq:rkcbsmumu}
C_{bs\mu\mu}=\sum_i\frac{\lambda_{\mu\mu ib}^*\lambda_{\mu\mu is}}{M_{\mu i}^2}\sim\frac{10^{-3}}{{\rm TeV}^2},
\eeq
into a constraint on the model parameters.
\begin{itemize}
\item $U_U$:
\beq
C_{bs\mu\mu}=\frac{|b|^2 y_\mu^2 y_t^2 V_{tb}V_{ts}^*}{M_{\mu t}^2}\ \Longrightarrow\ \frac{b}{M_U}\sim\frac{263}{{\rm TeV}}.
\eeq
\item $U_Q$:
\beq
C_{bs\mu\mu}=\frac{|b|^2 y_\mu^2 y_t^2 V_{tb}V_{ts}^*}{M_{\mu i}^2}\left(2{\cal R}e(x_{1/8})+y_t^2|V_{tb}|^2\right).
\eeq
In order that to have destructive interference with the SM amplitude, we need $2{\cal R}e(x_{1/8})+y_t^2|V_{tb}|^2>0$, namely (assuming that $x_{1/8}$ is real)
\beq\label{eq:rkuq}
x_{1/8}>-0.5.
\eeq
\item $U_D$:
\beq
C_{bs\mu\mu}=\frac{|b|^2 y_\mu^2 y_b^2 y_t^2 V_{tb}V_{ts}^*}{M_{\mu b}^2}\left(x_{1/8}^*+y_t^2|V_{tb}|^2\right).
\eeq
In order that to have destructive interference with the SM amplitude, we need $x_{1/8}+y_t^2|V_{tb}|^2>0$, namely
\beq\label{eq:rkud}
x_{1/8}>-1.
\eeq
\end{itemize}

\subsection{$b\bar b\to\tau^+\tau^-$}\label{sec:bbtautau}
Within the MFV models that we study, the requirement that the leptoquarks contribute to the Wilson coefficient of the operator of Eq. (\ref{eq:bsmumu}), namely to the $b\to s\mu^+\mu^-$ decay, implies that they contribute also to $b\bar b\to\ell^+\ell^-$ scattering processes \cite{Faroughy:2016osc}. MFV suggests that the largest contribution will be to the final $\tau^+\tau^-$ state. This contribution is constrained by the LHC searches for the $\tau^+\tau^-$ signature.

In Ref. \cite{Faroughy:2016osc}, the results of the ATLAS searches \cite{Aad:2015osa,Aaboud:2016cre} have been recast into bounds on vector leptoquarks mass and coupling:
\beq
C_{bb\tau\tau}^U=\frac{|\lambda_{\tau\tau bb}|^2}{M_U^2}\lsim2.6\ (4.0)\ {\rm TeV}^{-2},
\eeq
where the stronger (weaker) bound applies in case that $M_U>2$ TeV ($M_U\lsim2$ TeV), which is above (within) the LHC direct reach.  The bound for $M_U\lsim2$ TeV is not constant and is slightly weaker than 4 TeV$^{-2}$ below 1 TeV, which is anyway excluded by the LHC direct searches (see Appendix \ref{sec:directLHC}). In Fig. \ref{plot1} we present the excluded region for $M_U<2$ TeV, compared to the 1 $\sigma$ allowed region to fit the $R_K$ anomaly.

\begin{figure}[t]
 \begin{center}
   \includegraphics[width=0.59\textwidth]{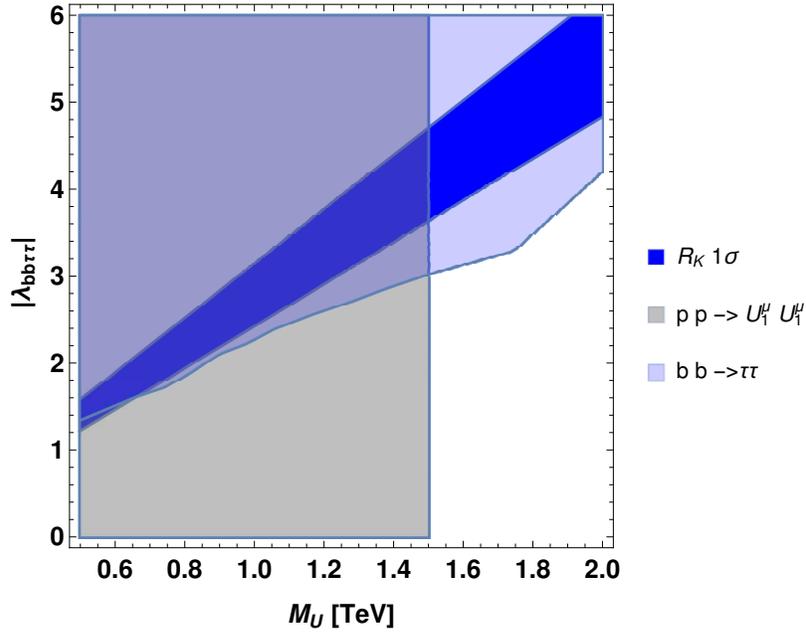} \
 \end{center}
    \caption{Constraints in the $M_U-|\lambda_{bb\tau\tau}|$ plane in the $U_U$ model: The $1\sigma$ allowed range from $R_{K^{(*)}}$, the region excluded by the high-$p_T$ $ pp\rightarrow\tau\tau$ search \cite{Faroughy:2016osc}, and the region excluded by LHC direct searches (see Appendix \ref{sec:directLHC}). }
\label{plot1}
\end{figure}

Within our models, we have [see Eq. (\ref{eq:rktautau})] $C_{bs\tau\tau}=0.28\ {\rm TeV}^{-2}$. We thus require
\beq\label{eq:bbttexp}
C_{bb\tau\tau}^U/C_{bs\tau\tau}^U\lsim9.3(14.3)\ {\rm for}\ M_U>(\lsim)2\ {\rm TeV}.
\eeq

The MFV prediction for $C_{bb\tau\tau}$ depends on the quark flavor representation:
\begin{itemize}
\item $U_U$:
\beq
\frac{C_{bb\tau\tau}^U}{C_{bs\tau\tau}^U}=\frac{V_{tb}}{V_{ts}}\sim25,
\eeq
which is excluded.
\item $U_Q$:
\beq\label{eq:bbbsuq}
\frac{C_{bb\tau\tau}^U}{C_{bs\tau\tau}^U}=\frac{(x_{1/8}+y_t^2|V_{tb}|^2)^2}{y_t^2 V_{tb}V_{ts}^*(2{\cal R}e(x_{1/8})+y_t^2|V_{tb}|^2)}.
\eeq
In the region allowed by $R_{K^{(*)}}$, $x_{1/8}>-0.5$ [see Eq. (\ref{eq:rkuq})], the function $(1+x_{1/8})^2/(1+2x_{1/8})$ has a minimum value of $1$, and consequently
\beq
\frac{C_{bb\tau\tau}^U}{C_{bs\tau\tau}^U}\gsim\frac{V_{tb}}{V_{ts}}\sim25,
\eeq
which is excluded.
\item $U_D$:
\beq
\frac{C_{bb\tau\tau}^U}{C_{bs\tau\tau}^U}=\frac{x_{1/8}+y_t^2|V_{tb}|^2}{y_t^2 V_{tb}V_{ts}^*}.
\eeq
Eqs. (\ref{eq:rkud}) and (\ref{eq:bbttexp}) imply a narrow allowed window:
\beq
-1<x_{1/8}\lsim-0.6(-0.4).
\eeq
\end{itemize}
We conclude that, within the MFV framework, the combined constraints from $R_{K^{(*)}}$ and $b\bar b\to\tau^+\tau^-$ exclude the $U_U$ and $U_Q$ scenarios, and leave the $U_D$ model as the only viable one.

\subsection{$R_{D^{(*)}}$}\label{sec:rd}
In addition to the measurements of $R_{K^{(*)}}$, there are hints of violation of LFU in $B\to D^{(*)}\tau\nu$ decay. Consider the ratios
\beq
R_{D^{(*)}}\equiv\frac{\Gamma(B\to D^{(*)}\tau\nu)}{\Gamma(B\to D^{(*)}\ell\nu)},
\ \ \ (\ell=e,\mu).
\eeq
The combination of measurements by BaBar \cite{Lees:2012xj,Lees:2013uzd}, Belle
\cite{Huschle:2015rga,Sato:2016svk,Abdesselam:2016xqt,Hirose:2016wfn} and LHCb \cite{Aaij:2015yra} reads \cite{Amhis:2016xyh}
\beq
R_D=0.403\pm0.047,\ \ \ R_{D^*}^{\rm SM}=0.310\pm0.017.\ \ \ \rho=-0.23.
\eeq
The SM predicts the following values \cite{Aoki:2016frl,Fajfer:2012vx,Bigi:2017jbd,Jaiswal:2017rve}:
\beq
R_D^{\rm SM}=0.300\pm0.008,\ \ \ R_{D^*}^{\rm SM}=0.252\pm0.003.
\eeq
Thus, there is a deviation from the SM prediction at $\sim4\sigma$, with $B\to D^{(*)}\tau\nu$ enhanced with respect to $B\to D^{(*)}\ell\nu$.

Within the MFV framework, the leptoquarks that generate the effective term of Eq. (\ref{eq:bsmumu}), relevant to $b\to s\mu\mu$ decays, generate also the term
\beq\label{eq:bctaunu}
C_{bc\tau\nu}(\overline{c_L}\gamma^\mu b_L)(\overline{\tau_L}\gamma_\mu\nu_{\tau L}).
\eeq
In contrast to the $b\to s\mu\mu$ and other processes discussed so far, the $b\to c\tau\nu_\tau$ decay is a quark-flavor changing {\it charged current} process. We have
\beq
C_{bc\tau\nu}^U=\sum_i\frac{\lambda^*_{\tau\tau ib}}{M_{\tau i}^2}
\sum_j \lambda_{\tau\tau ij}V_{cj}.
\eeq
The data require $C^{\rm NP}_{bc\tau\nu}\sim0.17\ {\rm TeV}^{-2}$. Together with the $R_{K^{(*)}}$ constraint, we need
\beq\label{eq:bctaubsmu}
C_{bc\tau\nu}^{\rm NP}/C_{bs\mu\mu}^{\rm NP}\sim-170.
\eeq

In the $U_D$ model, we have the following prediction for $C_{bc\tau\nu}$:
\beq
C_{bc\tau\nu}^{U}=\frac{|b|^2 x_{1/8}y_\tau^2 y_b^2 V_{cb}}{M_{\tau b}^2}
\left(x_{1/8}^*+y_t^2|V_{tb}|^2\right).
\eeq
Thus,
\beq
\frac{C^{U}_{bc\tau\nu}}{C^{U}_{bs\mu\mu}}=\frac{y_\tau^2}{y_\mu^2}\frac{V_{cb}}{V_{ts}^*}x_{1/8}\sim+(110-280),
\eeq
where the range corresponds to $-1<x_{1/8}\lsim-0.4$. The $U_D$ model predicts a strong {\it suppression} (by at least 10\%) of $R_{D^{(*)}}$ from the SM prediction and is thus excluded. In fact, it will remain excluded even if $R_{D^{(*)}}$ turns out to be consistent with the SM prediction (with experimental uncertainties no larger than the present ones) as long as $R_{K^{(*)}}$ is substantially suppressed compared to the SM.

We discuss additional aspects of $R_{D^{(*)}}$ within the MFV framework, independent of $R_{K^{(*)}}$, in Appendix \ref{sec:rtauk}.

\subsection{Summary of MQFV}
There are three classes of MQFV for leptoquarks that can generate the operator of Eq. (\ref{eq:bsmumu}):
\begin{itemize}
\item The $U_U$ model in the $(1,3,1)_{SU(3)_q^3}$ representation. It is excluded by a combination of the $R_{K^{(*)}}$ and $b\bar b\to\tau^+\tau^-$ measurements.
\item The $U_Q$ model in the $(3,1,1)_{SU(3)_q^3}$ representation. It is excluded by a combination of the $R_{K^{(*)}}$ and $b\bar b\to\tau^+\tau^-$ measurements.
\item The $U_D$ model in the $(1,1,3)_{SU(3)_q^3}$ representation. It is excluded by a combination of the $R_{K^{(*)}}$, $b\bar b\to\tau^+\tau^-$ and $R_{D^{(*)}}$ measurements.
\end{itemize}
We conclude that all MFV models considered by us are excluded.

\section{Conclusions}
\label{sec:conclusions}
The $R_{K^{(*)}}$ anomaly can be accounted for in models where there is a significant contribution to the $b\to s\mu^+\mu^-$ transition from the tree level exchange of leptoquarks. The pattern of deviations from lepton flavor universality (LFU) allows three simplified models, each with a single new leptoquark field:
\begin{itemize}
\item A Lorentz scalar, $SU(2)_L$ triplet: $T(3,3)_{-1/3}$;
\item A Lorentz vector, $SU(2)_L$ triplet: $U_3^\mu(3,3)_{+2/3}$;
\item A Lorentz vector, $SU(2)_L$ singlet: $U_1^\mu(3,1)_{+2/3}$.
\end{itemize}
Since the Yukawa couplings of these fields constitute new flavor parameters, they provide an opportunity to test various ideas for the flavor structure of new physics. In this work, we tested the idea of minimal flavor violation (MFV).

The need to break LFU, while keeping the $b\to s\tau^+\tau^-$ rates within bounds, implies that the new leptoquarks have to transform as $(1,\bar3)$ under the $SU(3)_L\times SU(3)_E$ lepton flavor group. On the other hand, without considering additional constraints, the representation under the quark flavor group $SU(3)_Q\times SU(3)_U\times SU(3)_D$ can be any of the three triplets, $(3,1,1)$, $(1,3,1)$ or $(1,1,3)$.

MFV relates the measured $B\to K^{(*)}\mu^+\mu^-$ rates to various other processes, such as $B\to K^{(*)}\nu\bar\nu$, $b\bar b\to\tau^+\tau^-$ and $b\to c\tau\nu$.  We summarize our use of these relations to test MFV in Table \ref{tab:exp}. Additional measurements ($B_s-\overline{B_s}$ mixing, direct leptoquark searches, $pp\to\mu^+\mu^-$, $\tau\to s\bar u\nu$) and considerations (perturbative unitarity) which are relevant to leptoquark models that aim to explain the $R_{K^{(*)}}$ anomaly, are discussed in Appendices.
\begin{table}[h!]
\caption{ \it MFV-predictions of simplified leptoquark models that account for the $R_{K^{(*)}}$ anomaly. $R_{\nu/\mu}$ is discussed in Section \ref{sec:rnumu}, $\Gamma_{bb\to\tau\tau}$ in Section \ref{sec:bbtautau}, and $R_{D^{(*)}}$ in Section \ref{sec:rd}. A super-index $*$ means that consistency with the observable applies for a small range of the parameter $x_{1/8}$.}
\label{tab:exp}
\begin{center}
\begin{tabular}{cccc} \hline\hline
\rule{0pt}{1.0em}%
Model & $R_{\nu/\mu}$ &
$\frac{\Gamma_{bb\to\tau\tau}}{\Gamma_{bb\to\tau\tau}^{\rm exp}}$ &
$R_{D^*}/R_{D^*}^{\rm SM}$ \\[2pt]
\hline\hline
Experiment & $<40$ & $<1$ & $1.23\pm0.07$ \\
\hline
$U_3^\mu(3,3)_{+2/3}$    & $5600$ && \rule{0pt}{1.0em}\\
$T(3,3)_{-1/3}$          & $310$  &&  \rule{0pt}{1.0em}\\
$U_{1U}^\mu(3,1)_{+2/3}$ & $8.5$  & $3.2$ &  \rule{0pt}{1.0em}\\
$U_{1Q}^\mu(3,1)_{+2/3}$ & $8.5$  & $\gsim3.2$ & \rule{0pt}{1.0em} \\
$U_{1D}^\mu(3,1)_{+2/3}$ & $8.5$  & $<1^*$ & $\lsim0.9$ \rule{0pt}{1.0em} \\
\hline\hline
\end{tabular}
\end{center}
\end{table}

Before we state our conclusions, let us repeat the ingredients of the models that we consider:
\begin{enumerate}
\item Simplified models, with a single leptoquark representation;
\item The leptoquark contribution to the $b\to s\mu^+\mu^-$ transition occurs at tree level;
\item The only spurion that breaks the lepton flavor symmetry is the charged lepton Yukawa matrix.
\item MFV is in the linear regime (higher powers in the spurions are suppressed compared to lower ones).
\end{enumerate}
Thus, we should bear in mind the following caveats:
\begin{enumerate}
\item Nature might have more than one leptoquark representation at play (see, {\it e.g.}, \cite{Dorsner:2017ufx});
\item Additional leptoquark representations might play a role via loop, rather than tree level contributions (see, {\it e.g.}, \cite{Bauer:2015knc,Becirevic:2016oho,Becirevic:2017jtw,Cai:2017wry});
\item Minimal lepton flavor violation might involve neutrino-related spurions (see, {\it e.g.}, \cite{Chiang:2017hlj});
\item The relations between third generation spurions and the lighter generation spurions are modified if MFV is in the nonlinear regime.
\end{enumerate}
Most of our conclusions hold, however, in generic such extensions of our framework. For example, even with tree level contribution to $B\to K^{(*)}\mu\mu$, the MFV framework predicts that the third generation couplings of the leptoquarks are close to the perturbative limit. If the contribution is suppressed by an additional loop factor, then these couplings will be pushed to non-perturbative values. As another example, if we allow neutrino-related spurions to play a significant role in lepton flavor conserving processes, it will be hard to avoid too large contributions to lepton flavor changing ones, such as $\mu\to e\gamma$ \cite{Dery:2013aba}.

Our conclusions do not hold, however, if MLFV is in the nonlinear regime. In this case, the strict relations between $\tau$ and $\mu$ couplings do not hold. Specifically, the bounds from $R_{\tau/\mu}$, $R_{\nu/\mu}$, $b\bar b\to\tau^+\tau^-$ and $B_s-\overline{B_s}$ mixing cannot be strictly applied. Yet, for some of the constraints, fine-tuned cancelations between the linear term and the higher order ones are needed to satisfy the constraints, which goes against the spirit of MFV. Order one modifications of the linear MFV prediction can, however, bring $U_1^\mu$ models into consistency with the $\Gamma_{bb\to\tau\tau}$ and $R_{D^*}$ constraints.
In fact, the phenomenology of models of nonlinear minimal flavor violation \cite{Kagan:2009bn} is similar to that of $U(2)$ models, which have been shown to be viable candidates to explain the $R_{K^{(*)}}$ anomaly \cite{Buttazzo:2017ixm}

We find that all models are excluded by a combination of $b\to s\mu^+\mu^-$, $b\to s\tau^+\tau^-$ and the processes presented in the Table. Note that for vector-leptoquark models, constraints from loop diagrams are sensitive to the UV completion of the model. It is thus important that we exclude these models based on tree level processes alone. (In Appendix \ref{sec:bsubs} we consider a loop process, $B_s-\overline{B_s}$ mixing, but we confront it with only scalar leptoquark models.)

We conclude that if the $R_{K^{(*)}}$ anomaly is experimentally established, then minimal flavor violation in the linear regime will be excluded.

\appendix

\section{$B_s-\overline{B_s}$ mixing}\label{sec:bsubs}
Leptoquarks which contribute at tree level to $b\to s\mu\mu$ contribute also via box diagrams to $B_s-\overline{B_s}$ mixing amplitude $M_{12}^s$. The modification of the SM prediction for $M_{12}^s$ is parameterized as follows:
\beq\label{eq:deltas}
M_{12}^s=|\Delta_s|e^{i\phi_s^\Delta}\times M_{12}^{{\rm SM},s}.
\eeq
Fitting the mixing amplitude to the experimental ranges of $\Delta m_{B_s}$, $\Delta\Gamma_{B_s}$ and $a_{\rm SL}^s$, gives \cite{Charles:2015gya}
\beq
|\Delta_s|=1.05^{+0.14}_{-0.13},\ \ \ \phi_s^\Delta=(1.5^{+2.3}_{-2.4})^o.
\eeq
Requiring that the contribution from the scalar leptoquark $T$ is within $\left||\Delta_s|-1\right|\lsim0.25$ gives \cite{Davidson:1993qk}:
\beq\label{eq:dmbs}
\frac{|\lambda_{\tau b}\lambda_{\tau s}|^2}{M_T^2}\lsim\left||\Delta_s|-1\right|\frac{192\pi^2\Delta m_{B_s}}{f_{B_s}^2 m_{B_s}}\sim2.0\times10^{-2}\ {\rm TeV}^{-2},
\eeq
where we used $\Delta m_{B_s}=1.17\times10^{-11}\ {\rm GeV}$, $m_{B_s}=5.37\ {\rm GeV}$ and $f_{B_s}\sim0.23\ {\rm GeV}$.

The $R_{K^{(*)}}$ anomaly requires (see {\it e.g.} \cite{DiLuzio:2017chi,DAmico:2017mtc})
\beq\label{eq:rkvalue}
C_{bs\mu\mu}=\frac{\lambda_{\mu b}\lambda_{\mu s}}{M_T^2}\sim10^{-3}\ {\rm TeV}^{-2}.
\eeq
In the viable models, $\lambda_{\tau b}^*\lambda_{\tau s}=(y_\tau/y_\mu)^2\lambda_{\mu b}^*\lambda_{\mu s}$, so that
\beq\label{eq:rktautau}
C_{bs\tau\tau}=\frac{\lambda_{\tau b}\lambda_{\tau s}}{M_T^2}\sim0.28\ {\rm TeV}^{-2}.
\eeq
Eqs. (\ref{eq:dmbs}) and (\ref{eq:rktautau}) can be simultaneously satisfied only for $M_T\lsim0.5$ TeV.

The members of the third generation $T$-triplet of charges $-4/3$, $-1/3$ and $+2/3$ decay into, respectively, $\tau b$, $\nu t$ and $\tau t$. The latter has branching ratio 1 which leads to an exclusion of 850 GeV \cite{Sirunyan:2017yrk}. The recast \cite{Diaz:2017lit} of the SUSY CMS analysis \cite{CMS:2017kmd} for the $t\bar t\nu\nu$ topology leads to an even stronger bound of 1.07 TeV. Thus, $M_T\lsim0.5$ TeV, as required by the $\Delta m_{B_s}$ constraint, is excluded by LHC direct searches. We conclude that $\Delta m_{B_s}$ constraints exclude the MFV-$T$ model as a possible explanation of the $R_{K^{(*)}}$ anomaly.

As concerns the case of vector leptoquarks, their contribution to $B_s-\overline{B_s}$ mixing is divergent. The divergence comes from the $k_\mu k_\nu$ term in their propagator, $i[(k_\mu k_\nu)/M_U^2-g_{\mu\nu}]/(k^2-M_U^2)$. Ref. \cite{Davidson:1993qk} suggests that a conservative bound can be obtained by considering the contribution of the $g_{\mu\nu}$ term only. The numerical factor of the mixing amplitude is four times larger than in the scalar case, and the resulting bound on the mass is therefore two times stronger, $M_U\lsim0.25$ TeV, which is excluded by the direct searches. Yet, this bound is model dependent. To relax the bound by a factor of ${\cal O}(7)$ (see Appendix \ref{sec:direct}), the contributions from the terms that we omitted should cancel with those that we took into account to the two percent level.

\section{LHC direct searches for $ U_1^{\mu}$}
\label{sec:directLHC}
The production cross section $\sigma(pp\rightarrow U_1^{\mu}U_1^{\mu\dagger})$ for vector leptoquarks is considered in \cite{DiLuzio:2017chi}. As already discussed in Section \ref{sec:mqfv}, MFV  implies that the nine flavor states are almost degenerate. This represents then an important and distinctive feature of our framework.

Each of the nine flavor-states has a branching ratio of 50\% to decay into a specific charged lepton and a jet, and a branching ratio of 50\% to decay into a neutrino plus jet. For each final state topology searched for at the LHC, we define
\beq
\sigma ( p p \rightarrow f f ) = \sigma ( p p \rightarrow U_1^{\mu}  U_1^{\mu\dagger})  N_f,
\eeq
with
\beq
N_f=\sum_{\alpha i}{\rm BR}(U_{\alpha i}\to f)^2.
\eeq
We ignore the mixed final states $ p p \rightarrow f f'$ depending on $\sigma ( p p \rightarrow U_1^{\mu}  U_1^{\mu\dagger})\sum_{\alpha i}{{\rm BR}(U_{\alpha i}\to f) \times {\rm BR}(U_{\alpha i}^\dagger\to f')}$.
In almost all cases, the decay is prompt. The only possible exceptions are the decays of $U_{eq}$, with $q=u$ or $d$, where the decay might have a displaced vertex.

The strongest bound come from the $U_1^\mu\rightarrow e b $ search \cite{ATLAS:2017hbw}:
\beq
\label{eb}
M_{e b}\gsim 1.5\ {\rm TeV}.
\eeq
The $U_{\tau b}$ state is the one related to both the perturbative unitarity bound, discussed in Appendix \ref{sec:direct}, and to the $ pp\rightarrow\tau\tau $ bound, discussed in Section \ref{sec:bbtautau}. It decays with branching ratios of $50\%$ into $\tau b$ and $50\%$ into $\nu t$, thus leading to $25\%$ of the events with $b\bar b\tau\tau$ final state and $25\%$ of the events with $\bar tt\nu\nu$ final state. These final states are constrained by, respectively, the search for third generation leptoquarks and a recast \cite{Diaz:2017lit} of the CMS SUSY search \cite{CMS:2017kmd}:
\beq\label{eq:mtaub}
M_{\tau b}\gsim1.0 \ {\rm TeV},\ \ \
M_{\nu t}\gsim 1.2 \ {\rm TeV}.
\eeq
(The latter bound is significantly stronger than the reach of the dedicated leptoquark search for the final state $ t \bar t \nu \nu$ \cite{Aad:2015caa} where the current limit is below 1 TeV.)

\section{Perturbative unitarity}
\label{sec:direct}
Perturbative unitarity requires for the  leptoquark vector-singlet case  of $U_1^\mu$ \cite{DiLuzio:2017chi}
\beq
|\lambda_{\tau\tau bb}|^2<4\pi.
\label{unpert}
\eeq
Requiring that the $R_{K^{(*)}}$ anomaly is accounted for by $U_1^\mu$ gives Eq. (\ref{eq:rkcbsmumu}). MFV relates $\sum_i\lambda_{\mu\mu ib}^*\lambda_{\mu\mu is}$ to $|\lambda_{\tau\tau bb}|^2$. Consequently, the combination of Eqs. (\ref{unpert}) and  (\ref{eq:rkcbsmumu}) leads to an upper bound on the leptoquark mass, in particular on $M_{\tau b}$, which can then be compared to the direct lower bound presented in Eq. (\ref{eb}) or, allowing for mass splitting within the $U_1^\mu$ multiplet, Eq. (\ref{eq:mtaub}).

\begin{itemize}
\item $ U_U$
\beq\label{eq:peruni}
|\lambda_{\tau\tau bb}|^2  \sim 7 \; \frac{M_{U_U}^2}{\rm TeV^{2}} < 4 \pi  \implies  M_{U_U}< 1.3 \;  \rm TeV,
\eeq
which is excluded by Eq.~(\ref{eb}) but not by Eq. (\ref{eq:mtaub}). Thus, a mass splitting larger than 200 GeV between $U_{\tau b}$ and $U_{eb}$ would be required to avoid the direct bounds and fulfill perturbative unitarity.
\item $ U_Q$
\beq
|\lambda_{\tau\tau bb}|^2  \sim \frac{7(1+x_{1/8})^2}{(1+2x_{1/8})} \frac{M_{U_Q}^2}{\rm TeV^{2}}  < 4 \pi \implies
M_{U_Q}  <\frac{1.3\sqrt{1+2 x_{1/8}} }{1+x_{1/8}} \rm TeV<1.3\ {\rm TeV},
\eeq
where, for the last inequality, we take into account that the correct sign of the $R_{K^{(*)}}$ anomaly requires $x_{1/8}>-0.5$. The situation is then similar to the $U_U$ case.
\item  $U_D$
\beq
|\lambda_{\tau\tau bb}|^2   \sim 7(1+x_{1/8})\frac{M_{U_D}^2}{\rm TeV^{2}}  < 4 \pi \implies M_{U_D} < \frac{1.3} {1+x_{1/8}} {\rm TeV},
\eeq
which is allowed by direct LHC searches for $ x_{1/8} < -0.24$.
\end{itemize}
We conclude that, assuming quasi-degeneracy within the $U_1^\mu$ multiplet, the combination of perturbative unitarity and LHC direct searches excludes the $U_U$ and $U_Q$ flavor models.

\section{$pp\to\mu^+\mu^-$}
\label{sec:ppmumu}
MFV models that contribute to $C_{bs\mu\mu}$ of Eq. (\ref{eq:bsmumu}), generate also the terms
\beq
C_{ss\mu\mu}(\overline{s_L}\gamma^\mu s_L)(\overline{\mu_L}\gamma_\mu \mu_L)+
C_{dd\mu\mu}(\overline{d_L}\gamma^\mu d_L)(\overline{\mu_L}\gamma_\mu \mu_L).
\eeq
These terms contribute to $pp\to\mu^+\mu^-$. Ref. \cite{Greljo:2017vvb} obtains from the experimental measurements the following bounds, which hold for leptoquark heavy enough that its effect on $pp\to\mu^+\mu^-$ is captured by EFT:
\beq\label{eq:ppmumu}
C_{dd\mu\mu}<0.023\ {\rm TeV}^{-2},\ \ \
C_{ss\mu\mu}<0.15\ {\rm TeV}^{-2}.
\eeq
Together with the $R_{K^{(*)}}$ constraints, these bounds imply
\beq\label{eq:ppmumu}
C^U_{dd\mu\mu}/C^U_{bs\mu\mu}\lsim23,\ \ \
C^U_{ss\mu\mu}/C^U_{bs\mu\mu}\lsim150.
\eeq

\begin{itemize}
\item $U_U$:
\beq
C_{dd\mu\mu}^U/C_{bs\mu\mu}^U=|V_{td}|^2/V_{ts}^*,\ \ \ C_{ss\mu\mu}^U/C_{bs\mu\mu}^U=V_{ts}.
\eeq
The $U_U$ contribution to $pp\to\mu\mu$ is negligibly small.
\item $U_Q$:
\beq
\frac{C_{dd\mu\mu}^U}{C_{bs\mu\mu}^U}=\frac{C_{ss\mu\mu}^U}{C_{bs\mu\mu}^U}=
\frac{|x_{1/8}|^2}{y_t^2V_{tb}V_{ts}^*\left(2{\cal R}e(x_{1/8})+y_t^2|V_{tb}|^2\right)}.
\eeq
Thus, for the $U_Q$ model, the $pp\to\mu\mu$ bound forbids $x_{1/8}$ outside the window $-0.4\lsim x_{1/8}\lsim+2.4$.
\item $U_D$:
\beq
\frac{C_{dd\mu\mu}^U}{C_{bs\mu\mu}^U}=\frac{|V_{tb}V_{td}|^2}{V_{tb}V_{ts}^*(x_{1/8}^*+y_t^2|V_{tb}|^2)},\ \ \
\frac{C_{ss\mu\mu}^U}{C_{bs\mu\mu}^U}=\frac{V_{tb}^*V_{ts}}{x_{1/8}^*+y_t^2|V_{tb}|^2}.
\eeq
Thus $U_D$ contribution to $pp\to\mu\mu$ is negligibly small, except for a small region which is excluded, $-1<x_{1/8}\lsim-0.99$.
\end{itemize}

\section{$R_{\tau/K}$ and $R_{D^{(*)}}$}
\label{sec:rtauk}
Here we discuss two tests of lepton flavor universality by charged current decays. We reconsider $R_{D^{(*)}}$, discussed above in Section \ref{sec:rd}, and we add another test, the $R_{\tau/K}$ ratio. We show that neither the $U_U$ model nor the $U_Q$ model can account for the $R_{D^{(*)}}$ anomaly, independently of the $R_{K^{(*)}}$ anomaly.

The $R_{\tau/K}$ ratio is defined via
\beq\label{eq:rtauk}
R_{\tau/K}\equiv\frac{\Gamma(\tau\to K\nu_\tau)}{\Gamma(K\to\mu\nu_\mu)}.
\eeq
Ref. \cite{Pich:2013lsa} translates the experimental measurements of the rates into the following bound (see their Table 2):
\beq
\frac{|C_{su\tau\nu}^{\rm SM}+C_{su\tau\nu}^{\rm NP}|}{|C_{su\mu\nu}^{\rm SM}+C_{su\mu\nu}^{\rm NP}|}=0.986\pm0.007,
\eeq
where $C_{su\ell\nu}$ is the Wilson coefficient of the term
\beq
C_{su\ell\nu}(\overline{u_L}\gamma^\mu s_L)(\overline{\ell_L}\gamma_\mu \nu_{\ell L}).
\eeq
The MFV models that we discuss contribute mainly to $C_{su\tau\nu}$:
\beq
C_{su\tau\nu}^U=\sum_i\frac{\lambda^*_{\tau\tau is}}{M_{\tau i}^2}
\sum_j \lambda_{\tau\tau ij}V_{uj}.
\eeq

We now obtain the predictions of the three $U_1^\mu$ MFV-models for $C_{su\tau\nu}$ and, where relevant, for $C_{bc\tau\nu}$.
\begin{itemize}
\item $U_U$:
\beq
C_{su\tau\nu}^U=C_{bc\tau\nu}^U=0.
\eeq
The $U_U$ model predicts
\beq
R_{\tau/K}=R_{\tau/K}^{\rm SM},\ \ \
R_{D^{(*)}}=R_{D^{(*)}}^{\rm SM}.
\eeq
Thus, the $U_U$ model cannot account for the $R_{D^{(*)}}$ anomaly {\it independently} of the $R_{K^{(*)}}$ anomaly.
\item $U_Q$:
\beq
C_{su\tau\nu}^U=\frac{|bx_{1/8}|^2y_\tau^2 V_{us}}{M_{\tau s}^2},\ \ \
C_{bc\tau\nu}^U=\frac{|bx_{1/8}|^2y_\tau^2 V_{cb}}{M_{\tau i}^2}.
\eeq
Thus
\beq
C^U_{su\tau\nu}/C^U_{bc\tau\nu}=V_{us}/V_{cb}=C^{\rm SM}_{su\tau\nu}/C^{\rm SM}_{bc\tau\nu},
\eeq
and the $U_Q$ model predicts
\beq
\frac{R_{\tau/K}}{R_{\tau/K}^{\rm SM}}=\frac{R_{D^{(*)}}}{R_{D^{(*)}}^{\rm SM}},
\eeq
which is strongly excluded.
\item $U_D$:
\beq
C_{su\tau\nu}=\frac{|b|^2x_{1/8}y_\tau^2 y_b^2 y_t^2 V_{ts}V_{tb}^*V_{ub}}{M_{\tau b}^2},
\eeq
which is negligibly small. The $U_D$ model predicts
\beq
R_{\tau/K}\approx R_{\tau/K}^{\rm SM}.
\eeq
The combination of the $R_{D^{(*)}}$ and $R_{K^{(*)}}$ constraints on $U_D$ was discussed in Section \ref{sec:rd}.
An analysis of the $U_D$ model with regard to $R_{D^{(*)}}$, independently of $R_{K^{(*)}}$, was carried out in Ref. \cite{Freytsis:2015qca}.

\end{itemize}
The analysis of $R_{\tau/\pi}\equiv\Gamma(\tau\to \pi\nu_\tau)/\Gamma(\pi\to\mu\overline{\nu_\mu})$ goes along similar lines.

\acknowledgments
YN is the Amos de-Shalit chair of theoretical physics.
YN is supported by grants from the Israel Science Foundation (grant
number 394/16), the United States-Israel Binational
Science Foundation (BSF), Jerusalem, Israel (grant number 2014230),
the I-CORE program of the Planning and Budgeting Committee and
the Israel Science Foundation (grant number 1937/12), and the Minerva Foundation.


\end{document}